# Quantum walks and signal transduction pathways

Agoni Valentina

valentina.agoni@unipv.it

**Signal transduction pathways recover a crucial role in cellular processes: they represent a connection between environmental conditions and cellular reactions. There are many pathways and they all are related to create a network. But how can every protein find the right way more fast than possible? How can it find the right down-streaming kinase in the cellular sea and not another very similar kinase?**
**Every signal transduction pathway can be seen as two distincted processes: the signal must reach**
**every kinase and then it must travel through the enzyme until its active site: quantum walks could be the answer to both the questions.**

Every signal transduction pathway is composed by one receptor and some kinases that bring the environmental signal to the nucleus. Usually, when the ligand binds the receptor, it activates a kinase by prosphorylation, the signal travel through the kinase and then the kinase activates the next one in the chain.

Electronic energy transfer involving oscillatory populations of donors and acceptors was first discussed more than 70 years ago[1].

Quantum coherence in photosynthetic complexes have been predicted [2,3] and indirectly observed[4].

Recently, direct evidence of long-lived coherence has been experimentally demonstrated for the dynamics of the Fenna-Matthews-Olson (FMO) protein complex[5].

Mohseni et al. (6) developed a theoretical framework for studying the role of quantum interference effects in energy transfer dynamics of molecular arrays.

However, the relevance of quantum dynamical processes to the exciton transfer efficiency is to a large extent unknown.

Quantum mechanics can explain the extreme efficiency, in that it allows the complexes to sample vast areas of space to find the most efficient path.

Many articles (7, 8, 9, 10) report electron transfer examples in polypeptides, but always refered to metalloproteins. However if we align the sequences of different kinases, we find that the most conserved residues are in the middle of loops and turns, as shown in figure 1 (11).

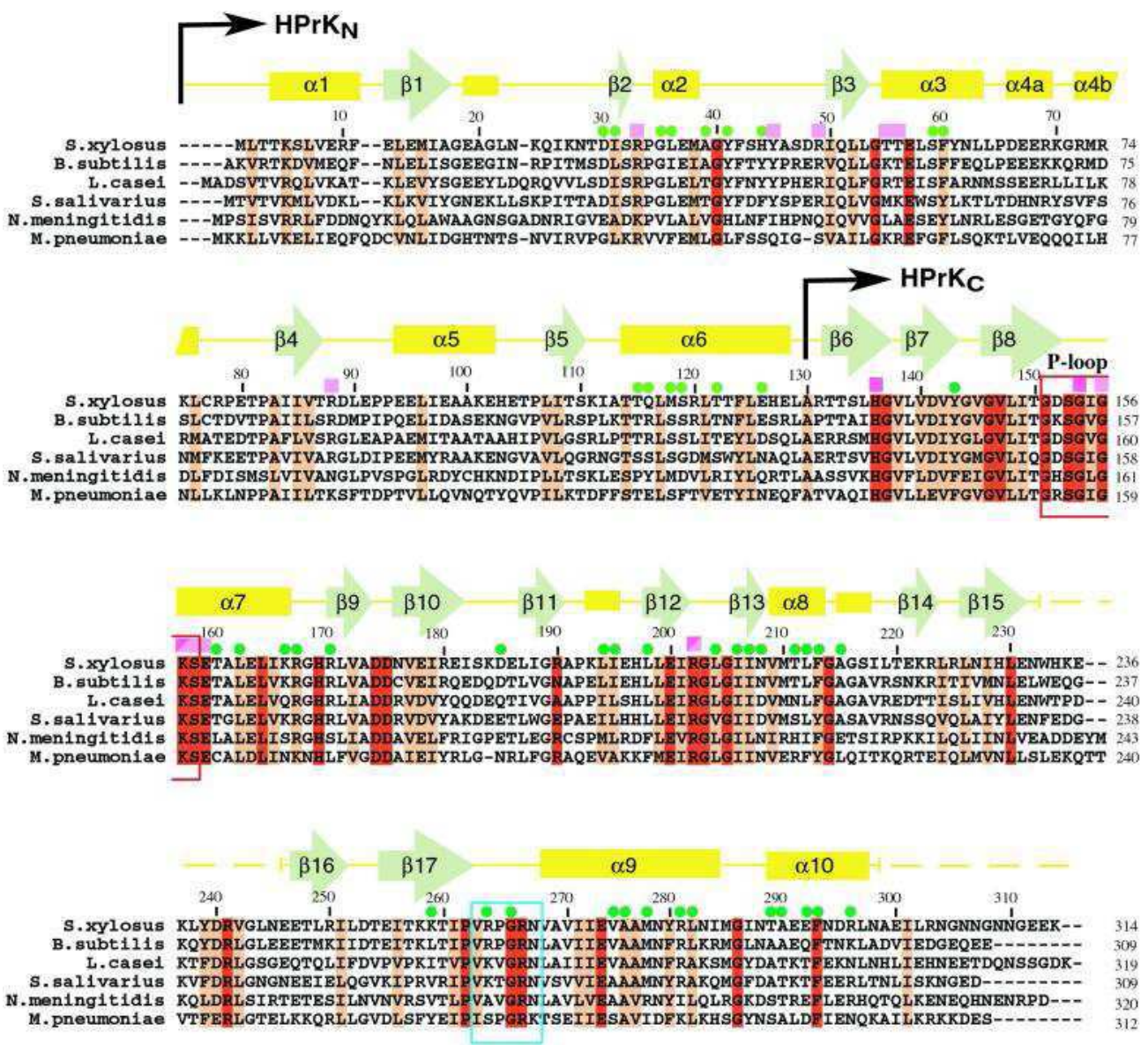

**Figure1.** Multiple sequence alignment of representative members of the HPrK protein family with the secondary structure assignment (11).

Moreover, considering kinases alignments at http://kinase.com/human/kinome/phylogeny.html positive and negative amino acids have an high frequency: 1/3 respect to 1/4 as a mean.
The large majority of protein kinases is activated by the phosphorylation of a polypeptide region (activation loop) that lies outside the active-site cleft. Analysis of the X-ray crystallographic structures of the insulin receptor with the activation loop in the phosphorylated and dephosphorylated forms offers a testable model for the mechanism of activity regulation by the loop[13].
In conclusion, these data indicate that inside the kinases the signal jumps (being an electron?) from a loop to another until it reaches the active site where phosphorylation takes place.
Figure 2 (14) represents the positive residues (in blue) that can reveal the path from the site of interaction with the previous enzyme, where phosphorylation takes place to the active site.

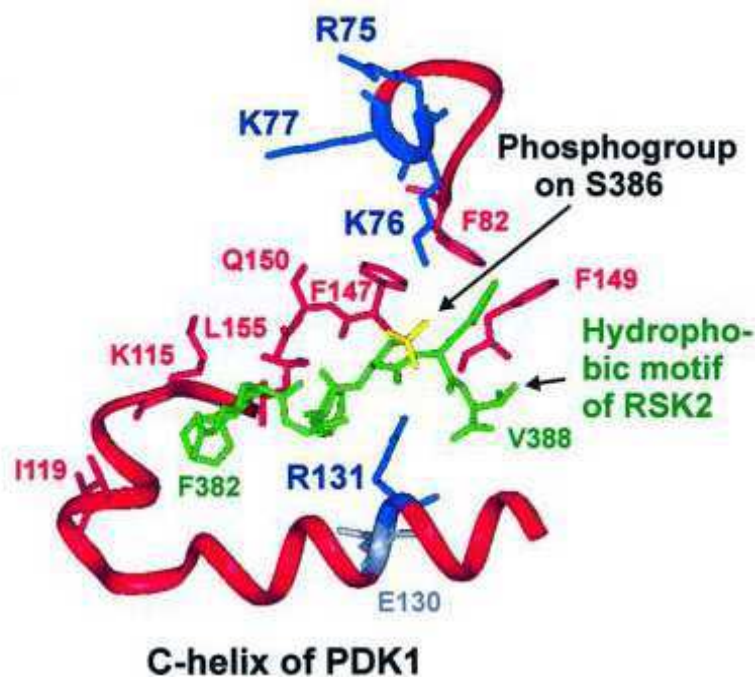

**Figure 2.** Model of the docking interaction between PDK1 and the phosphorylated hydrophobic motif of RSK2. PDK1 residues are red, blue or grey, whereas the RSK2 hydrophobic motif peptide is shown in green with the phosphogroup in yellow (14).

Every signal transduction pathway can be seen as two distincted processes: the signal travel through the kinase until its active site and then it must reach the next kinase.
Random walks (r.w.) allow to extimate the position of a particle after a time interval described by a probability distribution according to the nature of the process. Our goal is to check if a r.w. based model is useful to describe kinases' dynamics or if to consider quantum walks (q.w.). It is possible to calculate the time required to transport the signal to the nucleus, but this extimation do not take into consideration the probability of two kinases to match, equal to the probability of extracting two pre-extablished numbers over $15*10^8$. This calculation shows that we have to take into account q.w. instead of r.w.

In conclusion, quantum mechanics could lead to better understand signal transduction pathways involved in cancer and memory mechanisms and open the way for other quantum mechanical effects on biological systems.

1. Perrin, F. Thoérie quantique des transferts d'activation entre molécules de meˆme espèce. Cas des solutions fluorescentes. Ann. Phys. (Paris) 17, 283–314 (1932).
2. Knox, R. S. Electronic excitation transfer in the photosynthetic unit: Reflections on work of William Arnold. Photosynth. Res. 48, 35–39 (1996).
3. Leegwater, J. A. Coherent versus incoherent energy transfer and trapping in photosynthetic antenna complexes. J. Phys. Chem. 100, 14403–14409 (1996).

4. Savikhin, S., Buck, D. R. & Struve, W. S. Oscillating anisotropies in a bacteriochlorophyll protein: Evidence for quantum beating between exciton levels. Chem. Phys. 223, 303–312 (1997).

5. Engel, G.S., Calhoun, T.R.,. Read, E.L, Ahn, T.-K., Mancal, T., Cheng, Y-C., Blankenship, R.E. & Fleming, G.R. Evidence for wavelike energy transfer through quantum coherence in photosynthetic systems. Nature 446, 782 (2007).

6. Mohseni, M., Rebentrost, P., Lloyd, S. & Aspuru-Guzik, A. Environment-Assisted Quantum Walks in Photosynthetic Energy Transfer Journal of Chemical Physics 129, 174106 (2008).

7. Dolan, E. A., Yelle, R. B., Beck, B. W., Fischer, J. T. & Ichiye T.. Protein Control of Electron Transfer Rates via Polarization: Molecular Dynamics Studies of Rubredoxin. Biophysical Journal Volume 2030–2036 (2004).

8. Ferreira, A. M. & Bashford, D. Model for Proton Transport Coupled to Protein Conformational Change: Application to Proton Pumping in the Bacteriorhodopsin Photocycle J. Am. Chem. Soc. , 128 (51), 16778–16790 (2006).

9. Wang, P. & Blumberger, J. Mechanistic insight into the blocking of CO diffusion in [NiFe]-hydrogenase mutants through multiscale simulation. Proc. Natl. Acad. Sci. USA , 109, 6399 (2012).

10. Tipmanee, V. & Blumberger, J., "Kinetics of the terminal electron transfer step in cytochrome c oxidase" J. Phys. Chem. B 116, 1876 (2012).

11. Marquez, J.A., Hasenbein, S., Koch, B., Fieulaine, S., Nessler, S., Russell, R.B., Hengstenberg, W. & Scheffzek, K. Structure of the full-length HPr kinase_phosphatase from Staphylococcus xylosus at 1.95 Å resolution: Mimicking the product_substrate of the phospho transfer reactions. PNAS 99, 3458–3463 (2002).

12. http://kinase.com/human/kinome/phylogeny.html

13. Adams, J. A. Activation Loop Phosphorylation and Catalysis in Protein Kinases: Is There Functional Evidence for the Autoinhibitor Model? American Chemical Society 42 (2003).

14. Frödin, M., Antal, T. L., Dümmler, B. A., Jensen, C. J., Deak, M., Gammeltoft , S. & Biondi R. M. A phosphoserine/threonine-binding pocket in AGC kinases and PDK1 mediates activation by hydrophobic motif phosphorylation. The EMBO Journal 21, 5396-5407 (2002).